# Janus Lenses and Their Extraordinary Imaging Properties

Changbao Ma and Zhaowei Liu[*]

Department of Electrical and Computer Engineering, University of California, San Diego,
9500 Gilman Drive, La Jolla, CA 92093-0407
[*]zhaowei@ece.ucsd.edu

**Abstract:** Optical lenses are pervasive in various areas of sciences and technologies. It is well-known that the resolving power of a lens and thus optical systems is limited by the diffraction of light. Recently, various plasmonics and metamaterials based superlenses have been emerging to achieve super resolution. Here, we show that the phase compensated negative refraction lenses perform as "Janus Lenses", i.e. either converging lenses or diverging lenses depending on the illumination directions. Extraordinary imaging equations and properties that are different from those of all the existing optical lenses are also presented. These new imaging properties, along with the super resolving power, significantly expand the horizon of imaging optics and optical system design.

Lenses are the most fundamental and widely used elements in optics. Generally speaking, a lens is a refracting device that can reconfigure transmitted energy distributions. The configuration of a lens is determined by the type of wavefront conversion to perform; basic conversions include transformation among diverging waves, converging waves and plane waves. As any object can be considered as a group of point sources, a lens that converts a diverging wave to a converging one can form an image of the object.

Imaging is one of the most basic functions of a lens, of which the imaging properties have been well established and remain unchanged for centuries since the first elaboration by Kepler in 1611 [1]. As is summarized in Table 1 [2] and illustrated in Fig. 1, the image shows various characteristics when an object is placed at different locations with respect to the focal length $f$ of a lens. Such imaging rules, which are governed by the imaging equation $\frac{1}{s_0}+\frac{1}{s_i}=\frac{1}{f}$ with $s_0$ and $s_i$ being the object and image distances, respectively, contribute a bulky part of the ray optics and are the foundation of optical system design.

Based on the basic properties of a single lens, systems comprising multiple lenses can be conveniently designed. The resolution of an optical system is however beyond the scope of ray optics and has to be examined by wave optics. Owing to the diffractive nature of light, the best resolution of a lens system is limited to about half of the working wavelength $\sim\lambda/2$ [3]. Using the immersion technique, the resolution can be improved up to $\sim\lambda/(2n)$, but such enhancement however is quite modest due to the low refractive index *n* of natural materials [4, 5]. To achieve higher resolution, a material that can support the propagation of light with higher *k*-vectors is needed. Metamaterials, which are artificially engineered nanocomposites that can provide extraordinary material properties beyond those naturally available, provide new opportunities for this purpose [6-8]. Various metamaterial based superlenses that can achieve deep subwavelength scale resolution have been proposed and demonstrated within the last decade [9-21]. Despite their success in achieving super resolving power, such superlenses behave distinctively compared to their conventional counterpart. For instance, they cannot focus a plane wave, thus their imaging capability is exceedingly limited representing a significant defect from both theoretical and practical points of view.

By introducing various phase compensation mechanisms, the plane wave focusing function has been realized very recently [22, 23]. In this paper, we first provide a generalized imaging equation for the phase compensated metamaterials lenses, in which focal lengths can be defined from either direction. We further illustrate unprecedented imaging properties when such lenses also experience negative refraction at the lens/air interfaces.

For convenience purpose, we use the hyperbolic metalens, which is one type of the phase compensated negative refraction lenses, as a specific example to illustrate the new imaging paradigm. The process of designing such a hyperbolic metalens starts with focusing an incident plane wave from the air side, through phase compensation, to the focus $F_m$ at the focal length $f_m$ in the metamaterial, as shown in Fig. 2(a). The detail of the metalens design can be found in the Supplementary Online Material (SOM). Due to the hyperbolic dispersion of the metalens, a plane wave coming from the metamaterial side, instead of being focused, will diverge after the metalens, resulting in a virtual focus at $F_d$ also in the metamaterial (see Fig. 2(c)). It is commonly known that a conventional converging lens has two foci: one on each side; i.e., a plane wave coming from one side is focused to the other side, and vice versa. In stark contrast, a hyperbolic metalens works as a converging lens from one side but a diverging lens from the other side. Similar to a recent "Janus device" [24] demonstrating multiple optical functions along multiple directions based on the principles of transformation optics [25, 26], a hyperbolic metalens is a "Janus lens" having two different focusing behaviors (two faces) in opposite directions. Interestingly, although both focal lengths can be defined along forward and backward directions in the hyperbolic metalens, different signs must be assigned accordingly. This distinctive focusing behavior leads to extraordinary imaging properties that do not exist in common lenses.

With a given $f_m$ of a hyperbolic metalens, the imaging equation can be derived as (see detail in the SOM)

$$\frac{1}{v_d} + \frac{\varepsilon_z'/\sqrt{\varepsilon_x'}}{v_m} = \frac{\varepsilon_z'/\sqrt{\varepsilon_x'}}{f_m} \tag{1}$$

where $v_d$ and $v_m$ are the object/image distance in the dielectric (air) and metamaterial spaces, $\varepsilon_x$ and $\varepsilon_z$ are the permittivities of the metamaterial in the $x$ and $z$ directions, and the prime takes the real part. It can be seen from equation (1) that when $v_m \to \infty$, i.e., the incident beam is a plane wave from the metamaterial, the focal length in air is resulted as $f_d = f_m \sqrt{\varepsilon_x'}/\varepsilon_z'$. Replacing $f_d$ into equation (1), we obtain another form of metalens imaging equation

$$\frac{1}{v_d} + \frac{\varepsilon_z'/\sqrt{\varepsilon_x'}}{v_m} = \frac{1}{f_d} \qquad (2)$$

Because of the hyperbolic dispersion, $\varepsilon_z' < 0$, thus the relation between $f_d$ and $f_m$ indicates they always have different signs. According to the sign convention we define in the SOM, $f_m > 0$, so $f_d < 0$, meaning the $F_d$ of a hyperbolic metalens is not in air but in the metamaterial. Through an analogy to the typical symbolic notation of a focusing lens by using a double arrowed line, we symbolize the hyperbolic metalens with both foci in the metamaterial, as shown in Figs. 2(b) and (d).

As is analyzed above, a hyperbolic metalens breaks the forward/backward symmetry preserved in common lenses, leading to exceptional focusing behaviors. In the following, we will examine the imaging formation for an object placed at various locations with respect to its focal lengths. The specific hyperbolic metalens was designed with $f_m = 2.0$ μm at the wavelength of 690 nm. The metamaterial has a hyperbolic dispersion with $\varepsilon_x = 6.4 + 0.03i$ and $\varepsilon_z = -10.3 + 0.2i$.

Figure 3(a) schematically depicts the imaging behavior of such a hyperbolic metalens for an object in air. Two characteristic rays can be used to determine the image: the one parallel to the optical axis is "refracted" toward the focus $F_m$ in the metamaterial; another one towards the focus $F_d$ is "refracted" to the direction parallel to the optical axis in the metamaterial. The crossing point of these two rays determines the location of the image. Particularly, in the case of an object in air, the image in the metamaterial is always minified, erect, real and within the first focal length, regardless of the distance of the object in air to the metalens. This is completely different from a conventional lens, in which the image properties are dependent on the position of the object. Another special ray, which is the one passing the optical center $O$, may also be used to determine the location of the image through the calculation of its refraction angle in the metamaterial $\theta_m \approx \text{Re}(\sqrt{\varepsilon_x/\varepsilon_z})\sin\theta_d$, with $\theta_d$ being the incident angle and $\theta_m$ being the refracted angle of that ray with respect to the optical axis. Figure 3(b) shows the numerical verification for such a case using full wave simulation. Because the image can not go beyond the focal plane in the metamaterial space, the entire air space is mapped into the metamaterial space within the first focal plane.

Similarly, through schematic and geometric analysis, the image for an object in the metamaterial can be analyzed: the property of the image is dependent on the position of the object relative to the focus $f_m$. The image location can also be determined using characteristic rays: the one connecting the object and the focus $F_m$ is "refracted" to the direction parallel to the optical axis; another one parallel to the optical axis is "refracted" to the direction whose backward extension passes the focus $F_d$, because $F_d$ is not in air but in the metamaterial. Figure 4(a) shows the imaging behavior of an object within the first focal length in the metamaterial, i.e., $v_m < f_m$: the image in air is always magnified, erect and real. The image is real because the "refracted" rays converge in air. This case is simply the reciprocal of the case shown in Fig. 3. When the object is outside the first focal length, i.e., $v_m > f_m$ the rays in air diverge, but their backward extensions converge into a point on the other side of the optical axis in the metamaterial space, thus the image is always virtual and inverted. Depending on the relative position of the object, the image may be different in size; when the object is outside the first but inside the second focal length, i.e., $f_m < v_m < 2f_m$, the image is magnified, as shown in Fig. 4(c); when the object is at the second focal length, i.e., $v_m = 2f_m$, the image has the same size as the object, as shown in Fig. 4(e); when the object is outside the second focal length, i.e., $v_m > 2f_m$, the image is minified, as shown in Fig. 4(g). All of the four cases above for an object in the metamaterial are numerically verified, as shown in Figs. 4(b, d, f, h), respectively. These extraordinary properties of a hyperbolic metalens for an object in both air and the metamaterial are summarized in Table 2, which can also be directly derived from the metalens imaging equations (equation (1) and (2)).

Finally, as is shown in Figs. 3 and 4, the magnification of the metalens can also be determined analytically through the transverse magnification $M_T$ (more detail in the SOM)

$$M_T = \frac{u_m}{u_d} = -\frac{g_m}{f_m} = -\frac{f_d}{g_d} \qquad (3)$$

where $u_m$ and $u_d$ are the object/image height, i.e., the distance from the object/image to the optical axis, in the metamaterial and air space respectively; $g_m$ and $g_d$ are the object/image distance reckoned from their focus in the metamaterial and air space respectively.

Although illustrated using hyperbolic metalenses with $f_m > 0$, $f_m$ can also be designed to be negative ($f_m < 0$), meaning a plane wave from the air side diverges in the hyperbolically dispersive metamaterial. In this case, a plane wave from the metamaterial side will be focused in air, resulting in a positive $f_d$. Analyses show that all the above equations remain valid and the hyperbolic metalens with $f_m < 0$ is also a "Janus lens". The imaging properties of such a hyperbolic metalens with $f_m < 0$ are similar to those of the case of $f_m > 0$ but the air and metamaterial spaces are exchanged. (The imaging properties are summarized in the SOM)

We finally emphasize that the new set of imaging properties in the exampled hyperbolic metalens is a result of the hyperbolic dispersion of the metamaterials. The rules are applicable to any other type of phase compensated lenses that experiences negative refraction at the air/lens interface. These exotic imaging properties are impossible to realize by conventional lenses, thus extending the imaging properties and the capabilities of lenses, and the imaging optics in general, to a new horizon. The past centuries have illustrated the power of optical systems and devices based on conventional optical lenses in materials science, physics, chemistry, biology and even our daily life. The new found imaging behaviors and the super resolving power of hyperbolic metalenses provide new opportunities to explore novel optical devices and systems and will profoundly affect the advancement of optics.

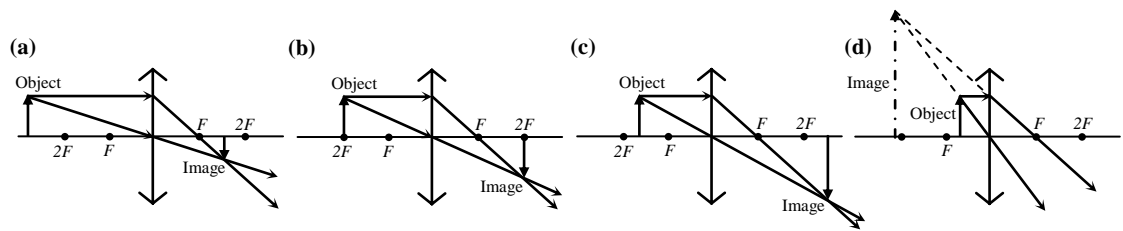

FIG. 1 (color online). Imaging behaviors of a conventional converging lens. (a) Object is outside point $2F$ at $2f$, where $F$ is the focus of the lens at its focal length $f$. (b) Object is at point $2F$. (c) Object is between points $F$ and $2F$. (d), Object is inside point $F$.

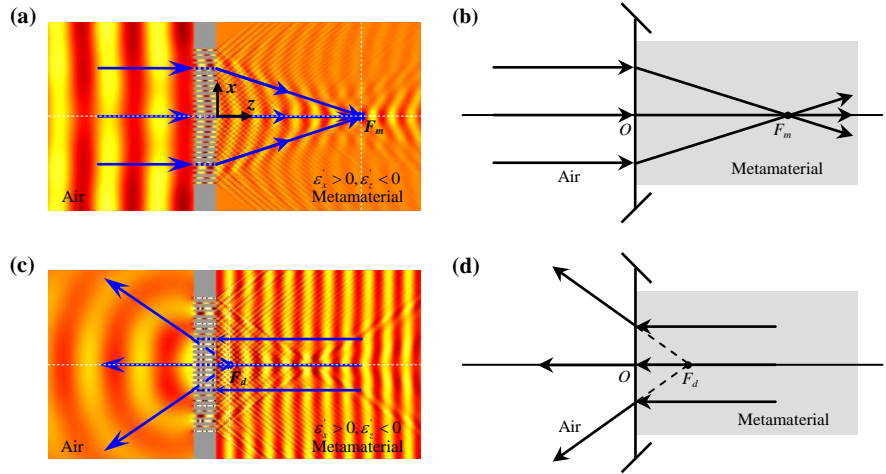

FIG. 2 (color online). Focusing behaviors of a hyperbolic metalens. (a) Plane wave from air converges to the focus ($F_m$) in the metamaterial. (b) Symbolized representation of the focusing behavior in (a). (c) Plane wave from the metamaterial diverges in air, resulting in a virtual focus ($F_d$) also in the metamaterial. (d) Symbolized representation of the diverging behavior in (c). The arrows in (a) and (c) are eye-guiding rays representing the directions of group velocities.

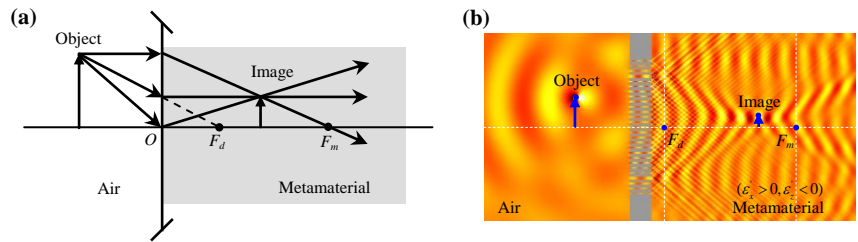

FIG. 3 (color online). Imaging behavior of a hyperbolic metalens for an object in the dielectric (air) space. (a) Schematic diagram. (b) Numerical verification.

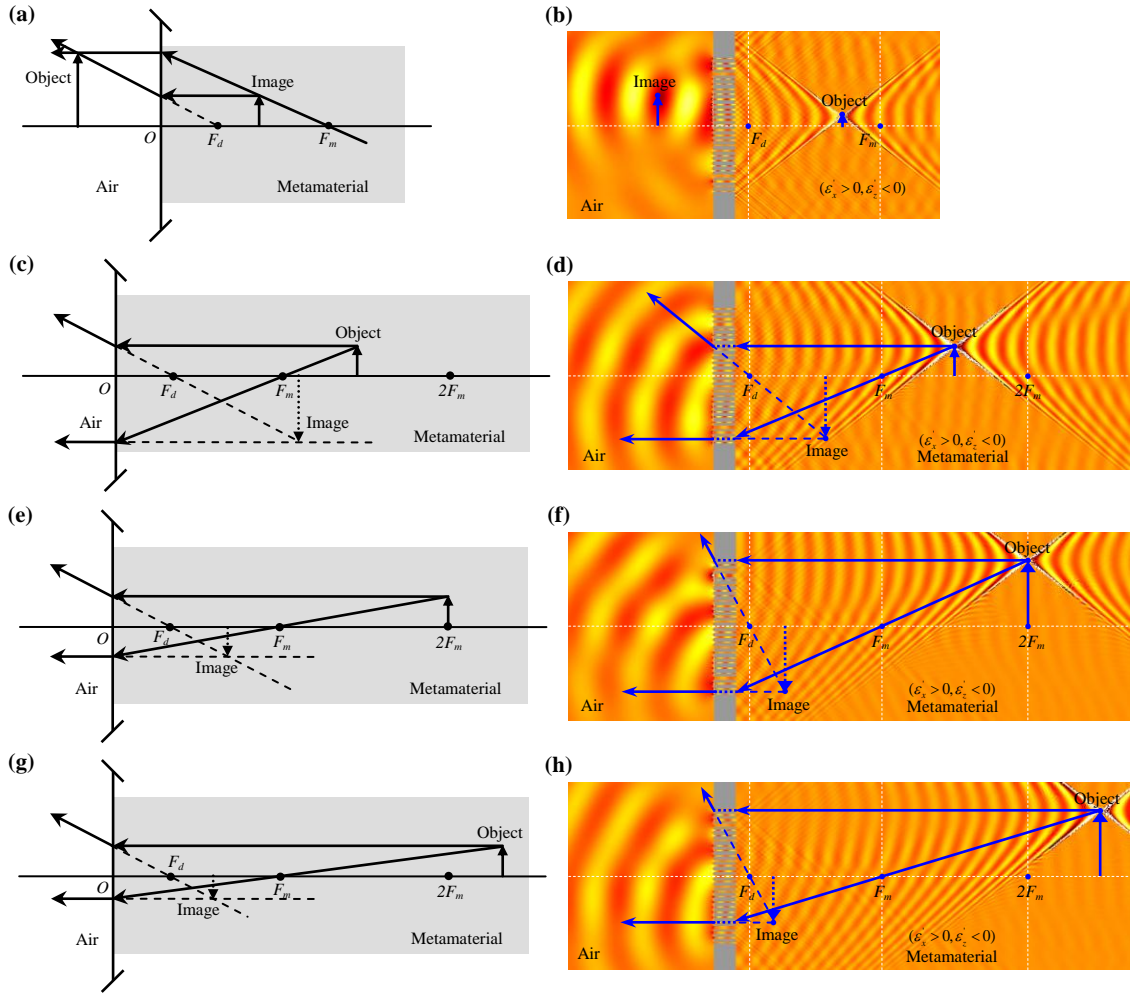

FIG. 4 (color online). Imaging behavior of a hyperbolic metalens for an object in the metamaterial space. (a) Object is inside point $F_m$. (c) Object is between points $F_m$ and $2F_m$. (e) Object is at point $2F_m$. (g) Object is outside point $2F_m$. (b, d, f, h) Numerical verification of (a, c, e, g) respectively.

| Object | Image | | | |
| --- | --- | --- | --- | --- |
| Location | Type | Location | Orientation | Relative size |
| $\infty>s>2f$ | Real | $f<p<2f$ | Inverted | Minified |
| $s=2f$ | Real | $p=2f$ | Inverted | Same size |
| $f<s<2f$ | Real | $2f<p<\infty$ | Inverted | Magnified |
| $s=f$ | | $\pm\infty$ | | |
| $s<f$ | Virtual | $-\infty<p<0$ | Erect | Magnified |

Table 1. Imaging properties of a conventional converging lens.

| Object | Image | | | |
|---|---|---|---|---|
| Location | Type | Location | Orientation | Relative size |
| $\infty > v_d > 0$ | Real | $0 < v_m < f_m$ | Erect | Minified |
| $\infty > v_m > 2f_m$ | Virtual | $2f_d < v_d < f_d$ | Inverted | Minified |
| $v_m = 2f_m$ | Virtual | $v_d = 2f_d$ | Inverted | Same size |
| $f_m < v_m < 2f_m$ | Virtual | $-\infty < v_d < 2f_d$ | Inverted | Magnified |
| $v_m = f_m$ | | $\pm\infty$ | | |
| $v_m < f_m$ | Real | $0 < v_d < \infty$ | Erect | Magnified |

Table 2. Imaging properties of a hyperbolic metalens. $f_m > 0$ and $f_d < 0$.

# Janus Lenses and Their Extraordinary Imaging Properties

## (Supplementary Material)

Changbao Ma and Zhaowei Liu[*]
Department of Electrical and Computer Engineering, University of California, San Diego,
La Jolla, California 92093-0407, USA
[*]Corresponding author zhaowei@ece.ucsd.edu

**I. Formulation for the metalens imaging equation, transverse magnification equation and the sign conventions**

A metalens, consisting of a metamaterial slab and a plasmonic waveguide coupler (PWC), is designed to have a focal length $f_m$ in the metamaterial using phase compensation for normal plane wave illumination from air. In analogy to the derivation of the imaging properties of a conventional optical lens, the imaging properties of a metalens can also be analyzed using the designed focal length $f_m$. Figure S1 shows the geometry and ray representation of a typical 2D metalens. An image $P_d$ in air may be formed for an object $P_m$ on the optical axis outside the focus $F_m$ in the metamaterial, and vice versa. The optical path length (*OPL*) from $P_d$ to $P_m$ is given by

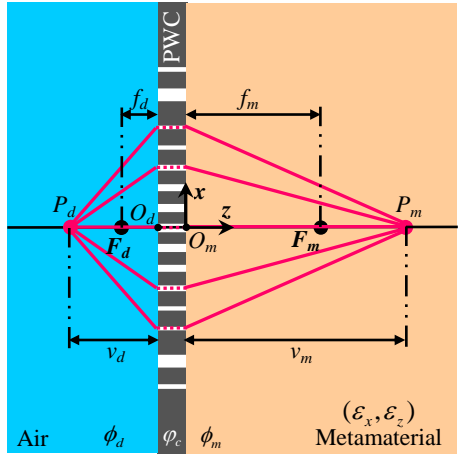

FIG. S1 (color online). Geometry of a metalens and the rays of two on-axis conjugate points. A metalens is designed to have a focal length $f_m$ in the metamaterial using plane wave incident from air.

$$[OPL] = (\phi_d + \phi_m + \phi_c)/k_0 \qquad (S1)$$

where $\phi_d = k_0\sqrt{v_d^2 + x^2}$ is the phase delay in air, $\phi_m = k_0\sqrt{\varepsilon_x' v_m^2 + \varepsilon_z' x^2}$ is the phase delay in the metamaterial, $\phi_c = \phi_{const} - k_0\sqrt{\varepsilon_x' f_m^2 + \varepsilon_z' x^2}$ is the designed phase delay profile of the PWC according to the metalens design principle [S1], with $\varepsilon_x'$ and $\varepsilon_z'$ being the real part of the electric permittivity of the metamaterial in the *x* and *z* directions respectively, $v_d$ being the image distance in air, $v_m$ being the object distance in the metamaterial and $\phi_{const}$ being a constant phase.

According to the Fermat's Principle, the optical path length remains stationary; that is, its derivative with respect to the position variable, which is $x$ here, is zero [S2], i.e., $d[OPL]/dx = 0$. So we have

$$\frac{n}{\sqrt{v_d^2 + x^2}} + \frac{\varepsilon_z'}{\sqrt{\varepsilon_x' v_m^2 + \varepsilon_z' x^2}} = \frac{\varepsilon_z'}{\sqrt{\varepsilon_x' f_m^2 + \varepsilon_z' x^2}} \tag{S2}$$

This relationship must hold true among the parameters for each ray from $P_m$ to $P_d$. Although this expression is exact, it is rather complicated due to the dependence on $x$. As in the formulation of the conventional glass lens, the paraxial approximation is assumed to simplify Eq. (S2). Thus with $n = 1$ for air, Eq. (S2) can be reduced to

$$\frac{1}{v_d} + \frac{\varepsilon_z'/\sqrt{\varepsilon_x'}}{v_m} = \frac{\varepsilon_z'/\sqrt{\varepsilon_x'}}{f_m} \tag{S3}$$

This is the metalens imaging equation, which is similar to the thin-lens equation [S2]. Note that in the reduction from Eq. (S2) to (S3), $v_d$, $v_m$ and $f_m$ are positive distances measured from $O_d$ and $O_m$ respectively; they can be negative when they are in the opposite side with respect to $O_d$ and $O_m$ respectively. This is the basis of the adopted sign convention for metalenses shown in Table S1. In Eq. (S3), when $v_d = \infty$, i.e., the metalens is illuminated with a plane wave in air, $v_m = f_m$, which is in consistence with the design. When $v_m = \infty$, we obtain the focal length in air

$$v_d = f_m\sqrt{\varepsilon_x'/\varepsilon_z'} = f_d = \overline{O_d F_d} \tag{S4}$$

Equation (S4) shows the relation between $f_d$ and $f_m$, with which Eq. (S3) can be rewritten as

$$\frac{1}{v_d} + \frac{\varepsilon_z'/\sqrt{\varepsilon_x'}}{v_m} = \frac{1}{f_d} \tag{S5}$$

The imaging equations (S3) and (S5), and the focal length in air $f_d$ of a metalens have been derived using the case in Fig. S1. As in a conventional optical glass lens, we adopt the following sign convention to generalize the case, as shown in Table S1. Note that $u_m$, $u_d$, $g_m$ and $g_d$ in Table S1 are geometric quantities and will be introduced later in Fig. S2.

| | | |
|---|---|---|
| $f_m$, $v_m$ | + | *Right of $O_m$* |
| $f_d$, $v_d$ | + | Left of $O_d$ |
| $u_m$, $u_d$ | + | Above axis |
| $g_m$ | + | Right of $F_m$ |
| $g_d$ | + | Left of $F_d$ |

Table S1. Sign convention for metalens.

According to this sign convention, $f_m$ is always positive as it is designed so in the metamaterial. Inasmuch as the metamaterial may have either elliptic ($\varepsilon_x' > 0$, $\varepsilon_z' > 0$) or hyperbolic ($\varepsilon_x' > 0$, $\varepsilon_z' < 0$) dispersion[S3], $f_d$ can be either positive for a metalens with an elliptically dispersive metamaterial, which is referred to as an elliptic metalens, or negative for a metalens with a hyperbolically dispersive metamaterial, which is referred to as a hyperbolic metalens. In the following, we show the derivation of Eq. (3) in the article.

We first assume an object is in air. Figure S2 shows the object $P_d(u_d, v_d)$ and image $P_m(u_m, v_m)$ locations for a hyperbolic metalens, where $u_d$ and $u_m$ are the distance of the object and image to the optical axis respectively and their signs are defined in Table S1. Because triangles $P_dAF_d$ and $BO_dF_d$ are similar and triangles $DO_mF_m$ and $P_mCF_m$ are similar, the Newtonian form of the lens equation can be easily obtained

$$f_m f_d = g_m g_d \tag{S6}$$

where $g_d$ and $g_m$ are the object and image distance measured from the focal point $F_d$ and $F_m$ in their space respectively. Their signs are defined in Table S1. Using Eq. (S6), the transverse magnification $M_T$ can be written as

$$M_T = \frac{u_m}{u_d} = -\frac{g_m}{f_m} = -\frac{f_d}{g_d} \tag{S7}$$

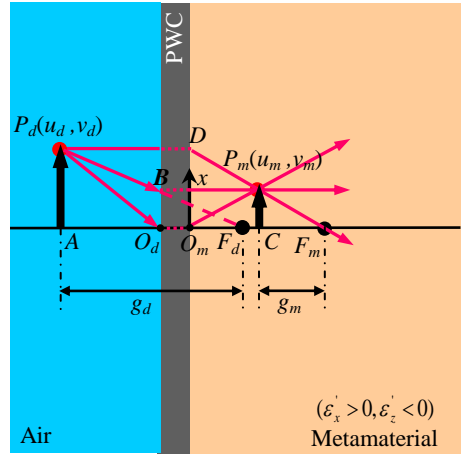

FIG. S2 (color online).   Object and image location for a hyperbolic metalens (object in air).

Equation (S7) is Eq. (3) in the article. These expressions in Eqs. (S6) and (S7) have the similar form as those for a conventional optical lens [S2]. The transverse magnification can be either positive for an erect image or negative for an inverted image. The meanings associated with the signs of the various parameters of a hyperbolic metalens are summarized in Table S2. It is worth noting that although Eqs. (S6) and (S7) are derived by assuming the object in air, they are generally applicable to all the cases analyzed in the article along with the adopted sign conventions.

| Parameter | Sign | |
|---|---|---|
| | + | - |
| $v_d$ | Real object | Virtual object |
| $v_m$ | Real image | Virtual image |
| $f_d$ | Converging | Diverging |
| $f_m$ | Converging | |
| $u_d$ | Erect object | Inverted object |
| $u_m$ | Erect image | Inverted image |
| $M_T$ | Erect image | Inverted image |

Table S2. Meanings associated with the signs of the metalens parameters.

## II. Imaging properties of a hyperbolic metalens with $f_m < 0$

| Object | Image | | | |
|---|---|---|---|---|
| Location | Type | Location | Orientation | Relative size |
| $\infty > v_m > 0$ | Real | $0 < v_d < f_d$ | Erect | Minified |
| $\infty > v_d > 2f_d$ | Virtual | $2f_m < v_m < f_m$ | Inverted | Minified |
| $v_d = 2f_d$ | Virtual | $v_m = 2f_m$ | Inverted | Same size |
| $f_d < v_d < 2f_d$ | Virtual | $-\infty < v_m < 2f_m$ | Inverted | Magnified |
| $v_d = f_d$ | | $\pm\infty$ | | |
| $v_d < f_d$ | Real | $0 < v_m < \infty$ | Erect | Magnified |

Table S3.   Images of a real object formed by a hyperbolic metalens. $f_m < 0$.

## III. Details of the simulated hyperbolic metalens and simulation parameters

All the simulations shown in the article are based on one hyperbolic metalens design at the wavelength of 690 nm. The metamaterial has a hyperbolic dispersion with $\varepsilon_x = 6.4 + 0.03i$ and $\varepsilon_z = -10.3 + 0.2i$, which is designed using silver nanowires in a dielectric background with a refractive index 1.3 at the silver volume filling ratio $p = 0.5$. The silver nanowires are orientated in the $z$ direction. The PWC is a nanoscale silver waveguide array which includes 27 nanoscale slits arranged symmetrically with respect to the optical axis. The waveguides are uniformly distributed with a center-to-center spacing of 70 nm but are designed to have different widths (from edge to center of the plate: 16.6, 21.9, 30.0, 43.9, 13.2, 15.0, 16.9, 19.0, 21.2, 23.3, 25.2, 26.8, 27.8, 28.2 nm) and thus provide

phase compensation for plane wave focusing to the metamaterial. The material in the slits has a refractive index 2.0 and the length of the waveguides is 300 nm. The permittivity of silver is -22.2 + 0.4i.

This hyperbolic metalens has a focal length in the metamaterial $f_m$ = 2.0 μm. Using Eq. S4, the calculated focal length for the air space is $f_d$ = -0.48 μm. These focal lengths are shown in Fig. 2 in the article.

Parameters for Fig. 3 in the article: The object in air is randomly chosen at ($u_d$ = 0.4 μm, $v_d$ = 0.7 μm), the minified real image is above $F_m$ at ($u_m$ = 0.118 μm, $v_m$ = 1.46 μm). Thus the transverse magnification is $M_T = u_m/u_d ≈ 0.3$ for this case, which is close to the calculated magnification 0.4 using Eq. S3.

Parameters for Fig. 4 in the article: Figure 4a is the reciprocal of Fig. 3, so the object is at ($u_m$ = 0.118 μm, $v_m$ = 1.46 μm) in the metamaterial. The object in Fig. 4d is at ($u_m$ = 0.4 μm, $v_m$ = 3.0 μm). The object in Fig. 4f is at ($u_m$ = 0.9 μm, $v_m$ = 4.0 μm). The object in Fig. 4h is at ($u_m$ = 0.9 μm, $v_m$ = 5.0 μm).